\documentclass[prl,aps,twocolumn,superscriptaddress,showpacs,floatfix,amssymb,amsmath]{revtex4-1}
\usepackage{epsfig}
\usepackage{graphics}
\usepackage{subcaption} 
\usepackage{float} 
\usepackage{array}
\usepackage{multirow}

\usepackage{lipsum}

\usepackage{changepage} 

\usepackage{setspace}

\usepackage{yfonts}

\usepackage{graphicx}
\usepackage{amssymb}
\usepackage{mathrsfs}
\usepackage{bbm}
\usepackage{epsfig}
\usepackage{makecell}

\usepackage[usenames,dvipsnames]{color}

\usepackage{bbm}
\usepackage{color}

\newcommand{\be}{\begin{eqnarray}}
\newcommand{\ee}{\end{eqnarray}}

\begin{document}

%

\title{
From Feynman's ratchet to timecrystalline molecular motors
}

\author{Jianmei Wang}
\affiliation{Center for Quantum Technology Research, Key Laboratory of Advanced Optoelectronic 
Quantum Architecture and Measurements (MOE), School of Physics, Beijing Institute of Technology, Beijing 100081, China
}
\author{Jin Dai}
\affiliation{Center for Quantum Technology Research, Key Laboratory of Advanced Optoelectronic 
Quantum Architecture and Measurements (MOE), School of Physics, Beijing Institute of Technology, Beijing 100081, China
}

\author{Antti J. Niemi $^\star$}
\affiliation{Nordita, Stockholm University and Uppsala University, Roslagstullsbacken 23, SE-106 91 Stockholm, Sweden }
%
\author{Xubiao Peng}
\affiliation{Center for Quantum Technology Research, Key Laboratory of Advanced Optoelectronic 
Quantum Architecture and Measurements (MOE), School of Physics, Beijing Institute of Technology, Beijing 100081, China
}


\begin{abstract}
Cats use the connection governing parallel transport in the space of shapes to land safely on their feet.
Here we argue that this connection also explains the impressive performance of molecular motors by enabling 
molecules to evade conclusions of Feynman's ratchet-and-pawl analysis. We first demonstrate, using simple molecular models, 
how directed rotational motion can emerge from shape changes even without angular momentum. We then
computationally design knotted polyalanine molecules and show how their shape space connection organizes 
individual atom thermal vibrations into collective rotational motion, independently of angular momentum. 
Our simulations show that rotational motion arises effortlessly even in ambient water, 
making the molecule an effective theory time crystal. Our findings have potential for practical molecular motor design and 
engineering and can be verified through high-precision nuclear magnetic resonance measurements.
  \end{abstract}

\maketitle
 
 \section*{Introduction}
 
Exploring the physical principles governing the operation of biomolecular machines, 
including proteins that are vital to all living cells,  is a challenging task yet has the 
potential to improve various aspects of human life \cite{thema-mm}.
However, developing synthetic and artificial molecular machines that can replicate 
their motion control and functionality remains a daunting undertaking \cite{Hanggi-2009, Kassem-2017, Hanggi-2002}.
The current understanding is that nonequilibrium 
statistical physics is the best approach  \cite{Brown-2020}.
This perspective originates from Feynman's exploration of the Brownian ratchet-and-pawl \cite{Feynman}. He 
showed that a ratchet in thermal equilibrium with its surroundings would only exhibit random thermal tumbling,
and the concept of a Brownian motor aims to 
overcome the limitations he identified \cite{thema-mm,Hanggi-2009, Kassem-2017, Hanggi-2002}. 
Subsequent investigations of linked and knotted molecular structures 
propose that topology also holds promise for advancing the field \cite{Gil-2015,horner-2016,Segawa-2019,vanraden-2019,Sawada-2019}.
 
Here we propose a novel perspective on how biomolecular motors generate and sustain 
directed rotational motion, even in the viscous environment of water at physiological temperatures. 
While Feynman's ratchet-and-pawl analysis posits that the ratchet relies on an external torque from 
the pawl to direct its rotation, we suggest that biomolecular motors are deformable bodies that utilize 
the geometry of their shape space to autonomously direct their thermal shape changes into a 
collective and sustainable rotational motion, without requiring any external torque or angular momentum. 
 Our novel paradigm builds on two theoretical ideas  that were 
both unknown to Feynman.  The first is the geometric concept 
of a connection in the shape space, 
originally  introduced by Guichardet \cite{Guichardet-1984},  Shapere and Wilczek
\cite{Shapere-1989a}. 
They explained why a deformable body can perform rotational motion simply by changing its shape. In mathematical terms, 
a continuous shape change is a trajectory in a space of all possible 
shapes, and the connection in this space relates a continuous shape change to parallel transport 
\cite{Montgomery-1993,Iwai,Littlejohn-1997,Marsden-1997}.
For cyclic shape deformations  the result is a 
rotational motion with a direction that depends on the connection.
The second idea  is the notion of a time crystal \cite{Wilczek-2012,Shapere-2012}
that explains how a physical object can rotate continually and effortlessly, even  in 
the lowest energy ground state of its thermodynamical  free energy.  In the case of a biomolecular motor, the ambient
water provides  a thermodynamically stable environment,  with recurrent  thermal collisions that continually  
change the shape of the molecule.   The connection \cite{Guichardet-1984,Shapere-1989a,Montgomery-1993,Iwai,Littlejohn-1997,Marsden-1997}
organizes these minute shape changes  into a directed, effective theory timecrystalline 
rotational motion  of the entire molecule that can persist indefinitely even in the absence of any angular momentum, 
until the system changes. 

 \section*{Theoretical considerations}
 
 We gain insight into our proposal, by examining the 
time evolution of a deformable triangle with three point-like interaction centers such as atoms or small molecules at the vertices
 $\mathbf r_a(t)$ ($a=1,2,3$); for clarity  we assign to each of them an equal unit mass.  Steric 
 effects prevent the vertices from overlapping, and with  no external forces the center of 
mass  is stationary and we  place it 
at the origin  $
\mathbf r_1 (t) +  \mathbf r_2 (t) + \mathbf r_3 (t) = 0
$. 
The shape can change arbitrary, provided the angular momentum vanishes
\vskip -0.3cm
\begin{equation}
\mathbf L \ = \ \mathbf r_1 \times \dot {\mathbf r}_1 + \mathbf r_2 
\times \dot {\mathbf r}_2 + \mathbf r_3 \times \dot {\mathbf r}_3 \ = \ 0
\label{Lz}
\end{equation}
The triangle then moves only on its own plane \cite{Guichardet-1984}  that we take to
coincide with the $z=0$ plane. 
Two shapes are the same when they differ at most  
by a rigid rotation around the $z$-axis.  To describe shape changes, 
we  assign to the vertices shape coordinates ${\mathbf s}_a = (s_{ax}, s_{ay})$. They describe  
all possible shapes of the triangle except a rotation,  when  we subject  them to  the following three conditions: 
First, the  vertex
${\mathbf s}_1(t)$  can move back and forth 
along the positive $x$-axis,  but  it can never leave this axis. 
Second, the  vertex ${\mathbf s}_2(t)$  can move freely on the upper half-plane
$s_{2y}(t)  >0$.  The position of the remaining  vertex ${\mathbf s}_3(t)$ is determined by our third 
condition ${\mathbf s}_3 (t) =  - {\mathbf s}_1(t) - { \mathbf s}_2(t)$.
The actual space coordinates $\mathbf r_a(t)$  can then deviate from the 
${\mathbf s}_a(t)$ at most by  an overall spatial rotation of the triangle around the $z$-axis 
\begin{equation}
\left( \begin{array}{c} r_{ax}(t) \\ r_{ay}(t) \end{array} \right) = 
\left( \begin{array}{cc}  \cos \theta(t)  &  - \sin \theta(t) \\  \sin \theta(t)  & \  \cos \theta(t) \end{array} \right) \left( \begin{array}{c} s_{ax}(t) \\ s_{ay}(t) \end{array} \right) 
\label{theta}
\end{equation}
and we  take $\theta(0)=0$ so that $\mathbf r_a(0) = {\mathbf s}_a(0)$ initially. 
To check whether there is any  rotational motion, 
we substitute (\ref{theta}) into   (\ref{Lz}).  
With
\[
I _{zz} = \sum\limits_{i=1}^3 \mathbf s^2_i
\ \ \ \ \& \ \ \  \mathcal L_z \ = \  \sum\limits_{a=1}^{3}  \left( s_{ay} \dot s_{ax}  - s_{ax} \dot s_{ay} \right)
\]
the $zz$-component of the moment of inertia tensor and
the $z$-component of the angular momentum in the shape space, respectively, the rotation angle 
at time $t$ is
\begin{equation}
\theta(t)  \ \equiv \ \int\limits_0^{t_1} \! dt^\prime \, \frac{d\theta(t^\prime)}{dt^\prime} \,  
\ = \  \int\limits_0^{t} \! dt^\prime  \,  I^{-1}_{zz} \mathcal L_z 
%
\label{dotheta}
\end{equation}
Guichardet,   
Shapere and Wilczek  \cite{Guichardet-1984,Shapere-1989a}  realized that the {\it r.h.s.} of (\ref{dotheta})  
defines a connection one-form
that is in  general non-vanishing. Its  integral along different shape space trajectories 
evaluates the rotational effect of different periodic shape deformations.

We have analyzed two simple molecule-inspired \cite{Wang-2021}
examples; additional examples can be found in \cite{Katz-2019,Peng-2021}. 
In both,  we 
describe shape deformations using   
bond lengths $D_{ab} =  |\mathbf s_a  - \mathbf s_b |$
that we combine into the distance matrix 
\begin{equation} 
M_{ab} \ = \ \frac{1}{2} ( D_{1b}^2 + D_{a1}^2 - D_{ab}^2 ) \ = \ (\mathbf s_a - \mathbf s_1) \cdot (\mathbf s_b - \mathbf s_1) 
\label{Mab}
\end{equation}
and we use Gram decomposition together with our three conditions  
to solve for  $\mathbf s_a(t)$, and  evaluate (\ref{dotheta}). 

\begin{figure}[h!]
 \begin{center}
    \includegraphics[width=0.45\textwidth]{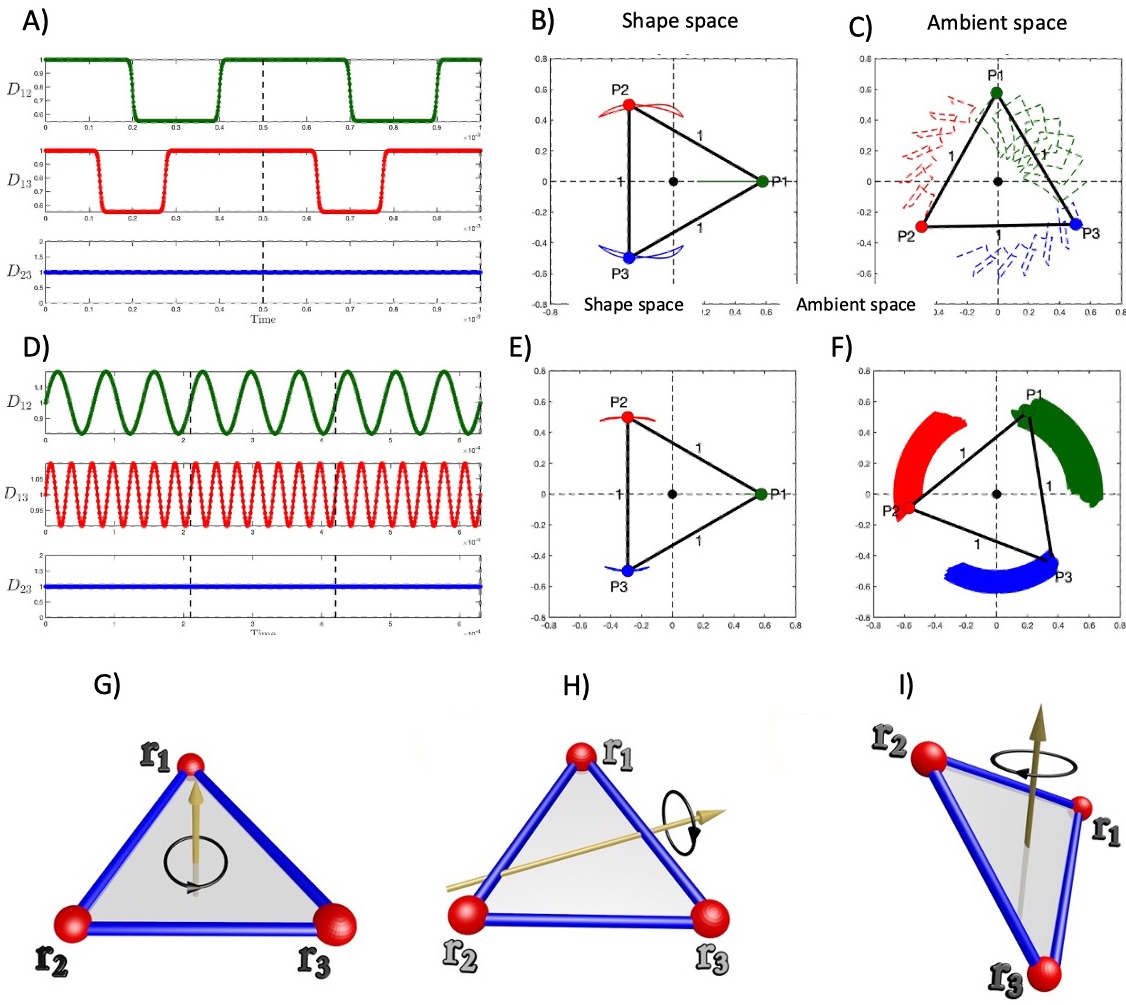}
      \end{center}
       \caption{
       Panel A depicts stepwise changes of bond lengths $D_{12}$ and $D_{13}$  between $D_{1a}=1.0$ and $D_{1a}=0.55$ 
       in an initially equilateral triangle, with $D_{23}=1$ fixed. Panel B  shows the motion in shape space, and panel C shows the actual  rotational 
       motion.
      Panels D-F show the same, in case of  harmonic bond length oscillations (\ref{hoL}) in an initially equilateral triangle $D_{ab}=1$ 
      for $t\in [0,0.1]$ 
      with  amplitudes $\Delta D_{12}=0.2$, $\Delta D_{13}=0.1$ and $T_1 = 7 \times 10^{-5}$ and $T_2 = 3 \times 10^{-5} $.
      Panel G and H show rotational motions with Hamiltonians (\ref{H3}) and  (\ref{H12}) respectively, and panel I depicts a
      generic rotational motion, as a combination of G and H.}
     \label{fig-1}
\end{figure}

In the first example we have an initially equilateral triangle that  changes its shape stepwise, with $D_{12}$
and $D_{13}$ shrinking and expanding cyclically while $D_{23}\equiv 1$ remains fixed as shown in panel A of figure 
\ref{fig-1}.   The panel B shows the motion in shape space, and panel C shows its rotational effect.
 The triangle oscillates back-and-forth and after six cycles we observe a $\sim 90^{\rm o}$ 
rotational motion. 
%
%
%
%
%
%
The second example describes  the rotational effect of bond length
oscillations when $D_{12}$ and $D_{13}$ evolve according to the  harmonic oscillator Lagrangian while $D_{23}=1$ is fixed.
\begin{equation}
L  =  \sum\limits_{a=2,3}\left\{  \frac{1 }{2}(\frac{ dD_{1a}}{dt})^2   -  \frac{1}{2} (\frac{2\pi}{T_a})^2  (D_{1a}-1)^2 \right\}  
 \label{hoL}
 \end{equation}
 At $t=0$ the triangle is equilateral. It
then oscillates around its symmetry axis, so that for generic 
parameter values we observe a net drift  
as exemplified in panels D-F of Figure \ref{fig-1}.

In both examples we observe a qualitative change in the dynamics in the limit of 
very small amplitudes and very high frequencies. This is the relevant limit for 
biomolecular applications, where the length and time scales are significantly larger than 
the individual atom oscillatory amplitudes and periods. The supplementary material's movie 1  
demonstrates how, in both examples, the limit is an equilateral triangle rotating uniformly
around its symmetry axis, as in panel G of figure \ref{fig-1}. 
This transition from a triangle with oscillating sites to a uniformly rotating equilateral triangle 
exemplifies the separation of scales, a general phenomenon that often 
enables the description of complex systems in terms of a few key variables and
commonly introduces qualitatively new features 
such as self-organization, collective oscillations, and emergent topological order.  To describe 
this limit in  our two examples,  we take  
the bond vectors $\mathbf n_i  = \mathbf r_{i+1} - \mathbf r_i$ 
with $\mathbf r_4 = \mathbf r_1$  as effective theory dynamical variables,  with SO(3) Lie-Poisson brackets \cite{Laurent}
and Hamiltonian 
\begin{equation}
H = e \, \mathbf n_1 \cdot \mathbf n_2 \times \mathbf n_3
\label{H3}
\end{equation}
Together with the equilateral triangle condition $\mathbf n_1 + \mathbf n_2 + \mathbf n_3=0$
Hamilton's equation for $\mathbf n_1$  is
\begin{equation}
\frac{d \mathbf n_1}{dt} = \{ \mathbf n_1, H \} = - \mathbf n_1 \times \frac{ \partial H }{\partial \mathbf n_1} =  
 \frac{e}{2} ( \mathbf n_{2} - \mathbf n_{3}) \ \not= \ 0
\label{n=3}
\end{equation}
with cyclic permutations  for $\mathbf n_2$, $\mathbf n_3$, and
the solution is a uniformly rotating equilateral triangle as in panel G of figure \ref{fig-1}.  
Note that  (\ref{H3}) vanishes for a triangle,  but the derivatives of $H$ do not vanish.
In fact, (\ref{n=3}) has no time independent solutions whatsoever, in the case of an
equilateral triangle. This qualifies it as an example of a  Hamiltonian time crystal  \cite{Dai-2019,Alekseev-2020},
an energy conserving  physical system that is in motion even at the minimum of  its mechanical free 
energy \cite{Wilczek-2012,Shapere-2012}.  
Apparently cyclopropane C$_3$H$_6$ represents this universality class \cite{Peng-2021}. 
The following effective theory Hamiltonian can also be introduced, in the case of a triangle.
\begin{equation}
H =  \mathbf n_1 \cdot \mathbf n_2 + g \mathbf n_2 \cdot \mathbf n_3
\label{H12}
\end{equation}
For $g\not=1$ it also describes a time crystal, with uniform rotational motion
around an axis that lies on the plane of the triangle and goes thru its center with a direction that depends
on the parameter $g$, as  shown in panel H of figure  \ref{fig-1}.
Finally, panel  I  depicts a time crystal with a
Hamiltonian that is a combination of (\ref{H3}) and (\ref{H12}).  With time dependent
$e(t)$ and $g(t)$ it can describe any rotational motion of an equilateral triangle around any axis  
that goes  through  the center of the triangle.

 For tangible molecular  examples, we designed polyalanine trefoil knots with 
varying lengths and studied the impact of their shape deformations using all-atom molecular 
dynamics simulations in both vacuum and water. We selected polyalanine as it is the simplest 
side-chained proteinomic amino acid. A closed chain was chosen to limit conformational 
entropy, making free energy minimization computationally feasible using available computers;
notably in the case of an open protein chain the free energy minimization remains a daunting task 
even for specialized supercomputers  \cite{Shaw-2021}. The choice of trefoil knot
topology for the chain was partially inspired by 
previous studies \cite{Gil-2015,horner-2016,Segawa-2019,vanraden-2019,Sawada-2019} that 
highlighted the potential significance of topology in the functionality of molecular motors. 
We also anticipate that the chirality of the trefoil knot aids in directing any rotational motion.

%
%
%

For simulations we have  utilized GROMACS \cite{gromacs} with the all-atom CHARMM36m force field \cite{charmm36m}. 
We have employed a three-step potential energy minimization protocol that we explain in detail in the Methods section. For our 
observations, following our triangular examples we have selected a reference triangle defined by three points that are located 
along the molecule's backbone, ideally symmetrically and at the positions of C$\alpha$ atoms. We connected these points using virtual segments 
and used the resulting triangle's time evolution to characterize the molecule's rotational motion. For this we used a quaternion representation
of rotations \cite{Hanson-book}:  
%
%
%
%
%
%
%
%
%
%
%
%
%
 \begin{figure}[h!]
  \begin{center}
    \includegraphics[width=0.45\textwidth]{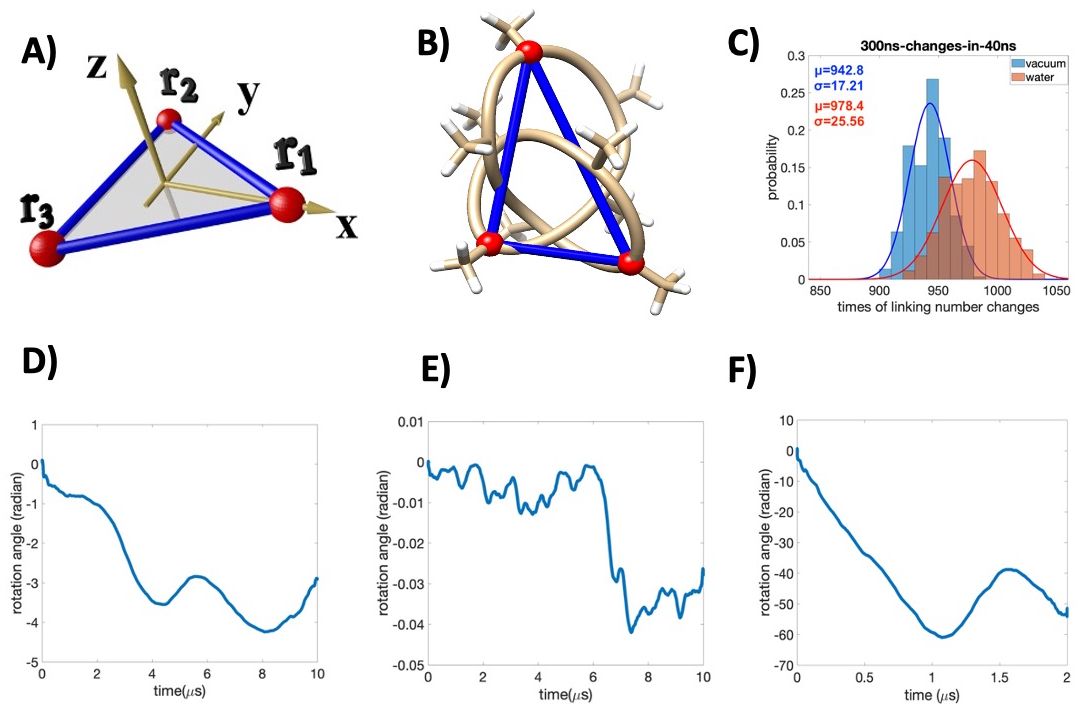}
      \end{center}
  \caption{Panel A shows the coordinate system and panel B is the minimum energy 9-ALA, together with the triangle we use to
  describe its rotational motion.  Panel C compares histograms for 9-ALA Gauss linking number changes between -2 and -3
    during a 40\,$n$s observation period,  in vacuum and in water.  Panel D shows 9-ALA 2\,$\mu$s moving average actual 
    evolution of $\vartheta_i$ in (\ref{quat}) in vacuum, panel E shows evolution of $\vartheta_i$ in vacuum when 
  evaluated from angular momentum using
  (\ref{rho}) and panel F shows 0.4\,$\mu$s  moving average actual evolution of  $\vartheta_i$ in water.} 
\label{fig-2}
\end{figure} 
As shown in panel A of Figure \ref{fig-2}, at each time step $t_i$ we introduced a cartesian coordinate system, with 
origin at the center of mass of the entire molecule, and we adjusted the reference triangle so that its geometric center coincided with the molecule's
center of mass. We then assigned to the axes an orthonormal basis of 
unit quaternions ($\mathbf{\hat  i}_i, \mathbf{\hat  j}_i, \mathbf{\hat  k}_i$) with
$\mathbf{\hat  i}_i$ pointing from the center of mass to one of the 
vertices  of the triangle ($\mathbf r_1$ in the figure), with $\mathbf{\hat  k}_i$ the normal to the plane of the triangle, and with
$\mathbf{\hat  j}_i$ determined by right-handed orthonormality.  At time step $t_i$ the instantaneous Euler axis of rotation 
determines a unit quaternion $
\mathbf N_i  = x_{i}  \mathbf{\hat  i}_i + y_{i}  \mathbf{\hat  j}_i + z_{i}  \mathbf {\hat k}_i $ where $ x_{i}^2 + y_{i}^2 + z_{i}^2 = 1$. 
The unit normal $\mathbf k_i$ at time $t_i$ is then related to the initial
normal vector $\mathbf k_1$ by 
\begin{equation} 
  \mathbf k_i  = \exp\{ - \frac{\vartheta_i}{2} \mathbf N_i \}
 \, \mathbf k_1 \,  
\exp\{  \frac{\vartheta_i}{2} \mathbf N_i \}
\label{quat}
 \end{equation}
where $\vartheta_i$ is the rotation angle around the Euler axis, and we use it  to  characterize 
the rotational motion of the triangle 
relative to the initial triangle:  In the case of a rotational motion on a plane $\vartheta_i$ coincides with the rotation
angle $\theta$ of our triangular examples.


In our  simulations the initial molecule  had vanishing angular momentum, but due to 
round-up error accumulation,  and Brownian rotational tumbling when water-molecule 
interactions are present, the angular momentum can fluctuate
during the simulation.
Thus  we monitored its value, to  isolate  its contribution from the rotational motion of the molecule
that is due to  shape deformations. 
For this, imagine that between time steps  $t_i$ and $t_{i+1}$ the molecule rotates as a rigid body. To describe this
"rigid body" {\it i.e.} proper rotation of the molecule, we evaluated
 the components of the instantaneous moment of inertia tensor $\mathbb I_{ab}(i)$   and the components $ L_a(i)$
of the instantaneous angular momentum. We then evaluated 
the components of the instantaneous "rigid body" angular velocity
$ 
\omega_{a}(i) = \mathbb I^{-1}_{ab} (i) L_b(i)
$. 
With $\mathbf r_i $  is the position  of a generic atom with respect to the molecule's center of mass at time $t_i$,
its  putative "rigid body" rotated position at $t_{i+1}$ is then
\begin{equation} 
 \mathbf r_{i+1} \ = \    \mathbf r_i  + 
 ( t_{i+1}-t_i) \,  \boldsymbol \omega_i \times  \mathbf r_i 
\label{rho}
\end{equation}
 The difference between the  actual orientation and the putative orientation evaluated according to (\ref{rho}) is then 
the rotational motion that we  attributed to the molecule's shape  changes;
we found that  the rotation
evaluated using (\ref{rho}) is quite small in comparison to the rotational motion by shape changes.
We note that the "rigid body" rotation (\ref{rho}) is tantamount  to a description using
instantaneous Eckart frames \cite{Eckart-1935} where a rotational motion due to shape 
changes can not be detected. 
%
%
%
%
%
%
%
%
%
%
%


 \section*{Results}
 
 We now describe our simulation results for 9 amino acid polyalanine trefoil  which is the shortest trefoil knot that we can  design
without any steric conflicts, and 42-ALA with  is the shortest trefoil knot that we can design
with all peptide planes in the {\it trans}-conformation at energy minimum. 
In the case of 9-ALA our  minimum energy all-atom structure 
together with the triangle that we used to follow its rotational motion, 
is shown in Panel B of figure \ref{fig-2}.   The peptide planes are neither 
in {\it trans} nor in {\it cis} conformation implying that  the backbone is strained. Notably,
the energy minimum breaks spontaneously the three-fold $D_3$  trefoil symmetry 
into a $C_2$ symmetry with  three  distinct, degenerate
minimum energy trefoils that are related to each other by $i\to i+3$ $mod(9)$  shift along the backbone. 
In the case of 42-ALA 
the $D_3$ trefoil symmetry is similarly spontaneously 
broken into $C_2$, with the three energy minima related to each
other by a $i\to i+14$ $mod(42)$ shift. 

Our vacuum  trajectories were 10\,$\mu$s long at constant 310\,K internal molecular  temperature; note that 
thermal effects could be thought of as rudimental approximation of quantum mechanical zero point fluctuations.  
In the case of 9-ALA the average  C$\alpha$ backbone root-mean-square 
distance (RMSD)  between the initial structure and those along the trajectory was stable at
0.14\,$\pm$0.02\,\AA, and for 42-ALA we obtained RMSD 0.69\,$\pm$0.15\,\AA.
The $D_3$ symmetries remained broken in both cases.
We used  {\it Topoly} \cite{Dabrowski-2021} to evaluate the evolution of Gauss linking number between 
the C$\alpha$  trace and the virtual C$\beta$ trace, which in the present case
is a {\it local} topological invariant. For 9-ALA the initial value was -2 and  along the trajectory
it oscillated frequently  between -2 and -3,  as shown in panel C of figure \ref{fig-2}. 
 For 42-ALA the initial  value was -1  and along the trajectory it oscillated between 0 and  -2.

\begin{figure}[h!]
  \begin{center}
    \includegraphics[width=0.45\textwidth]{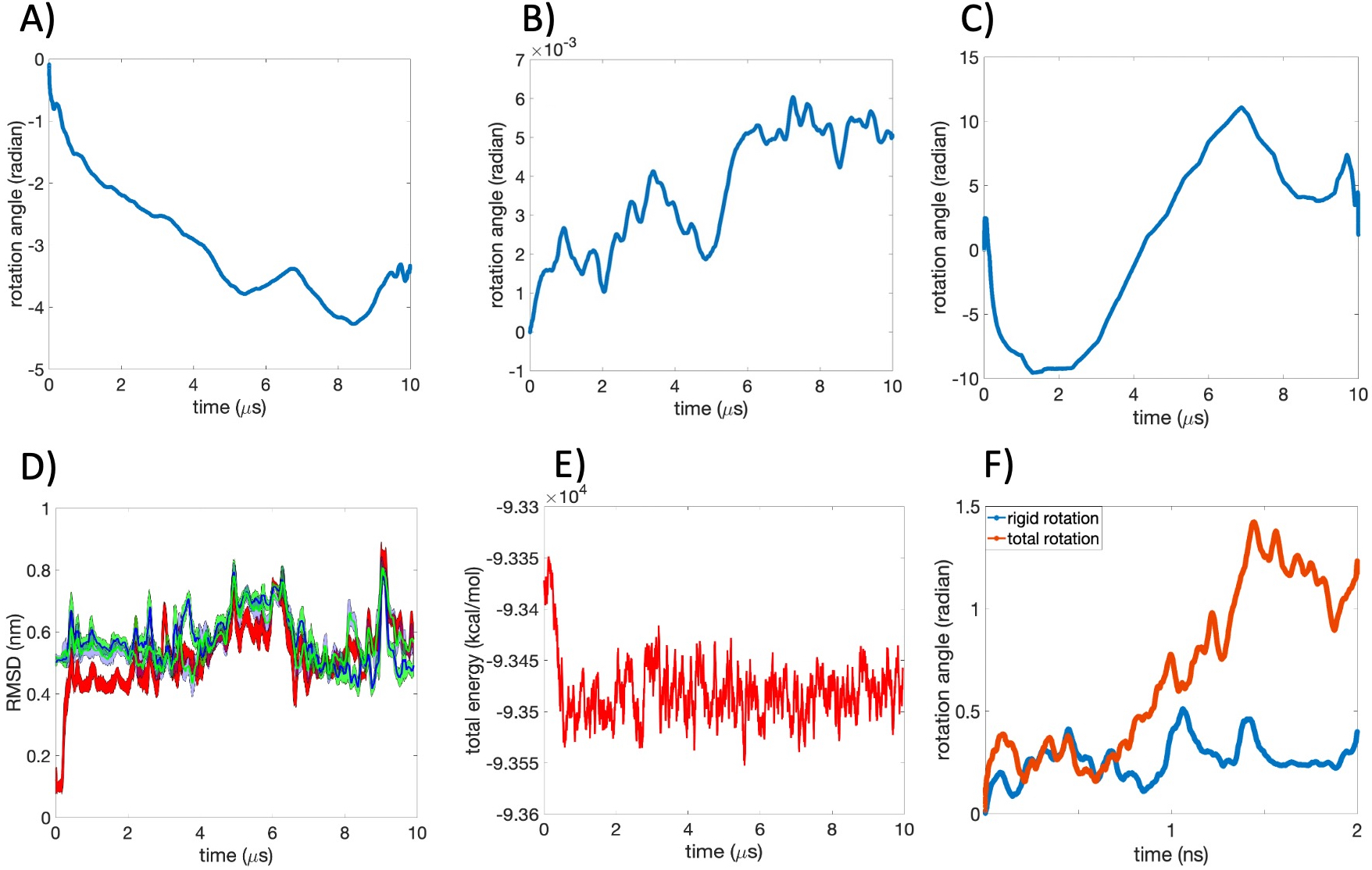}
      \end{center}
  \caption{Panel A shows rotational motion of 42-ALA in vacuum in terms of $\vartheta_i$ in (\ref{quat}). 
  Panel B shows the $\vartheta_i$ evolution
  in vacuum, when evaluated from instantaneous angular momentum. Panel C shows evolution of $\vartheta_i$ in water. Panel D shows RMSD from 42-ALA 
  trajectory in water to initial energy minimum (red),
  and to the other two $\mathbb Z_3$ symmetric energy minima (blue and green). Panel E shows total energy along water trajectory of 42-ALA. Panel F shows 
  a generic short segment of the water trajectory,  comparing the actual rotational motion to that evaluated from (\ref{rho}). The difference is  the 
  rotational motion due to shape deformations.
} 
\label{fig-3}
\end{figure}

Panel D of figure \ref{fig-2}  shows the  2\,$\mu$s moving average  time evolution of the  rotation 
angle $\vartheta_i$ in  (\ref{quat}), in the case of 9-ALA.  Panel A of figure \ref{fig-3} shows the same for 42-ALA.
In both, we have  back-and-forth oscillations in combination of an overall  directed rotational motion,
 in resemblance of panels C and F in figure \ref{fig-1}. 
The panel E of figure \ref{fig-2} shows the evolution of  $\vartheta_i$ when evaluated from 
(\ref{rho}) for 9-ALA, and B of figure \ref{fig-3} shows the same for 42-ALA.
Clearly, the contribution from  angular momentum fluctuations to the rotational motion due to shape deformations
was vanishingly small, in both cases. 


%

We  have simulated the 9-ALA  trefoil  extensively in a solvent with 1157 explicit water
molecules at 310\,K, and we find that  the molecule retains its (average) shape with no indication
of $D_3$  symmetry restoration: The C$\alpha$ backbone RMSD between the initial structure 
and those along the trajectory is  0.13\,$\pm$0.02\,\AA.  But the 
water-molecule interactions  intensify the frequency of shape fluctuations that drive the molecule's rotational motion.
This can be seen  in panel C of figure \ref{fig-2} that shows how the Gauss linking number oscillates between -2 and -3 
and a comparison between panels D and F shows how the rotational motion is clearly 
 faster in water.  But it becomes difficult to compare the actual rotational motion with the instantaneous 
 rigid body rotation computed from (\ref{rho}). This is because water exchanges angular momentum with the molecule, 
 and  in a box with periodic boundary conditions it is problematic to ensure conservation of total angular momentum.


%
%
%
%
%
%
%
%
%

We have embedded the minimum energy 42-ALA vacuum structure  in solvent  with 2445 explicit 
water molecules, and minimized the total energy by gradient descent.
The length of our 310\,K dynamical production run is 10\,$\mu$s. Initially, $\vartheta_i$ decreases. 
But after 200-300\,$n$s  there is a rapid transition where the direction of rotational motion appears to reverse (panel C of figure-\ref{fig-3}),
the RMSD between the initial structure and those along the trajectory increases to $\sim$4.5\,\AA,
the RMSD from the trajectory to the two other $C_2$ symmetric energy minima approach the same value (panel D),
and  the  Gauss linking number of the C$\alpha$-C$\beta$ traces changes from -1 $\pm 1$  to -9 $\pm$2. 
The total energy also decreases, as shown in panel E:  We conclude that the spontaneously broken $D_3$ symmetry becomes restored, 
in a manner that resembles a second order phase transition akin the ferromagnetic to paramagnetic  phase transition in magnetic materials. 
Panel F in figure \ref{fig-3} compares the observed rotational motion to that computed from (\ref{rho}). There
is a clear distinction between the two.

In the limit of large stroboscopic time steps where an effective theory description becomes valid,
the dynamics both in the case of 9-ALA and 42-ALA  
resembles a tumbling time crystal 
with the evolution of the reference triangle described by a Hamiltonian  that
is a sum of (\ref{H3}) and (\ref{H12}) with time dependent parameters $e(t)$ and $g(t)$

 \section*{Conclusions}

Our high-precision all-atom molecular dynamics study shows that the concept of a connection can be essential to the design and control 
of effective molecular machines. Our results demonstrate how  biomolecules can  convert individual atom thermal vibrations into a rotational 
motion of the entire molecule by utilizing the connection in their shape space. This  challenges the prevailing view of molecular motors as 
rigid bodies by highlighting the crucial role of deformability in their functionality. In particular,
we have  discovered that the impressive effectiveness of many biomolecular motors can be modeled by an effective theory
Hamiltonian time crystal while at the atomic level the  dynamics is driven by the heat bath 
of ambient water molecules.  We have also exposed the dynamical consequences of a spontaneous symmetry breakdown and restoration. 
Finally, although simulations of open protein chains are computationally much more demanding  we expect our 
findings to hold true for such systems as well.

%
%


%

\section*{Acknowledgements}

A. J. N. is supported by the Carl Trygger Foundation Grant No. CTS 18:276, by the Swedish Research Council under Contract No. 2018-04411, and by COST Action CA17139. 
The work by J.W. and X.P. are supported by Beijing Institute of Technology Research Fund Program for Young Scholars.  Nordita is supported in part by Nordforsk.

\vskip 0.8cm
\subsection*{Additional information}

%

\vskip 0.3cm
\noindent
{\bf Data and code availability:}  Raw data and analysis codes are available from the corresponding author.

\vskip 1.5cm

\noindent
\section*{Supplementary material:}  

The supplementary material consists of two movies, and description of simulation methods

\vskip 0.2cm
\noindent
$\bullet~$ The first movie in file {\bf movie1.mp4 } describes  the qualitative change in dynamics in the case of the 
model Lagrangian  defined in equation (\ref{hoL}) in the text. The movie starts with
parameter values that describe an oscillating triangle, with no observable rotational motion.
Then the parameters including  time scale change stepwise. As a consequence the frequency of oscillations increase and
their amplitudes decrease and eventually become 
vanishingly small. At the same time the rotational
motion becomes more apparent, until the triangle appears to rotate uniformly in line with the scaling limit effective theory timecrystalline
Hamiltonian (\ref{H3}), even though the angular momentum vanishes. 

\vskip 0.2cm
\noindent
$\bullet~$
The second movie in file {\bf movie2.mp4}  displays a segment of the trajectory of the 42-ALA in water at 310\,K, as described in the text. 
It shows the first 1.0\,$\mu$s with 10\,$n$s between frames, starting from the initial condition and 
covering the $C_2 \to D_3$ symmetry restoring transition. The rotational motion is clear from the orientation of the reference triangle, and
the structural change during the symmetry restoration can be seen by 
comparing the first and last frames. 

\vskip 1.0cm

\subsection*{Simulation Methods}

In our all-atom polyalanine trefoil knot simulations we use GROMACS \cite{gromacs} with the all-atom CHARMM36m force field \cite{charmm36m}.  
We first minimize the free energy, using a three-step protocol. For this we 
start with vacuum, with  free boundary conditions  and both Coulomb and van der Waals interactions 
extending over all atom pairs in the molecule with no cut-off approximation.

\vskip 0.1cm
\noindent
$\bullet$ In step one we  start with the trefoil template
\begin{equation*}
\begin{split}
&x(s)  = (2 + \cos 3s) \cos 2s 
\\
& y(s) =(2 + \cos 3s) \sin 2s  \ \ \ \ \ \ \ \ \ s\in [0,2\pi] \\
& z(s) = \sin 3s
\end{split}
\end{equation*}
We discretize this into a linear polygonal chain
with $n$ equidistant vertices;  $n=9$ and 42 in othe examples we describe in the text.
The neighboring vertices are  3.8\,\AA \ apart, which is the average distance 
between neighboring C$\alpha$ atoms along a protein backbone. 
We use  PULCHRA  \cite{Rotkiewicz-2008} to construct  an initial
all-atom representation.

\noindent
$\bullet$ Step two minimizes the energy using a kinetic diffusion process. This
consists of  an iterative  series of energy conserving double precision all-atom  runs.
For each  run we choose  the initial kinetic energies of all the atoms to be zero. The typical length
of a single run is 1.0\,$p$s, during which some of the excess potential energy becomes 
converted into kinetic energy of the atoms, that we remove to start the next iteration. We repeat the process 
until we observe no kinetic energy, and minimize the final potential energy using LBFG \cite{Byrd-1995}.


\noindent
$\bullet$ Step three aims to remove the above structure from a putative local energy minimum towards a global minimum, using
double precision annealing. We start with a 10\,$n$s heating phase from 0\,K to 500\,K followed by a 5\,$n$s stabilization.
We then proceed to a 985\,$n$s cooling simulation that brings  the temperature back to 0\,K.

\vskip 0.2cm

We repeat the steps two and three until we observe no energy decrease in the  final  0\,K configuration. This proposes 
that we have reached the minimum of the molecule's potential energy.  In the case of 9-ALA our minimization algorithm decreases
the CHARMM36m energy from the initial value 25583 kJ/mol down to 6120 kJ/mol  and in the case of 42-ALA
the CHARMM36m energy goes down from 6255 kJ/mol to 1764 kJ/mol.

We then employ  energy drift, a common phenomenon in all-atom simulations, to construct the corresponding  
finite temperature minimum energy 
configuration:  We initiate a single precision run with 1.0\,$f$s time step.  Due to error accumulation 
the internal temperature of the molecule starts  increasing, and we let the energy drift proceed 
until the temperature has reached a target value,  310\,K in the examples that we present here.  
This gives us the initial configuration of  our production runs in vacuum.

A tight knotted protein such as our polyalanine trefoils,  is subject to
steric restraints proposing  that the minimum energy configurations in vacuum and in water should be
geometrically close to each other. Thus, we start all our water simulations  by placing the above 310\,K minimum 
energy molecule at the center of a cubic box with periodic boundary condition,  with 12\,\AA ~ distance  to the 
edges. We fill the box with explicit TIP3P \cite{TIP3P} water,  with 12\,\AA \  cutoff for Coulomb and  van 
der Waals interactions. We then minimize the potential energy of the entire system, using gradient descent to construct a 
minimum energy ensemble. We bring the entire system 
in thermal equilibrium at 310\,K, and we start the production run always  with vanishing initial angular momentum. 
We keep the simulation time 
step short enough to eliminate detectable energy  drift, and  we  use double precision as need be.

%

\end{document}